# Dynamic Modulation of Electronic and Optical Properties in GaN Bilayers by Interlayer Sliding

Heeju Kim[1] and Gunn Kim[1*]

[1]Department of Physics & Astronomy and Hybrid Materials Research Center, Sejong University, Seoul 05006, Republic of Korea

*Corresponding author: gunnkim@sejong.ac.kr

## Abstract

In this study, we present a first-principles investigation of the electronic and optical properties of gallium nitride (GaN) bilayers, focusing on the influence of interlayer sliding and spacing. In contrast to the earlier studies on discrete stacking configurations, we explore the dynamic evolution of the properties during transitions between stable stacking arrangements. Using density functional theory calculations, we systematically analyze the impact of these structural variations on the electronic band structure and optical absorption spectra of GaN bilayers. The analysis includes both high-symmetry stacking configurations (AA', AB', and AC') and intermediate states generated by controlled in-plane atomic displacements, thereby providing a comprehensive understanding of the property changes associated with interlayer sliding. The findings of this study provide valuable insights into the potential for tuning the electronic and optical response of two-dimensional GaN for applications in nanoscale photonic and electronic devices, where precise control over interlayer interactions and stacking is crucial.

# 1. Introduction

Gallium nitride (GaN) is a wide bandgap semiconductor with a direct bandgap of approximately 3.4 electronvolts (eV) [1, 2]. This material has had a profound effect on the fields of optoelectronics and high-power electronics. Its superior properties compared to silicon, including efficient light emission and high breakdown voltage, have enabled breakthroughs in blue/ultraviolet light-emitting diodes (LEDs) [3, 4], laser diodes (LDs) [5, 6], and high-electron-mobility transistors (HEMTs) [7-9]. The applications include solid-state lighting [10, 11], power conversion [12, 13], and 5G/6G communication [14, 15]. Typically crystallizing in the wurtzite structure, GaN exhibits inherent anisotropy, giving rise to piezoelectric [16] and polar characteristics [17] that are crucial for various device functionalities.

The advent of two-dimensional (2D) materials, exemplified by graphene [18], hexagonal boron nitride (h-BN) [19-21], and transition metal dichalcogenides (TMDs) [22-25], has opened unprecedented opportunities for tailoring material properties through quantum confinement and surface effects. This dimensional reduction has inspired intense research on 2D GaN [26-28], aiming to harness its unique potential for next-generation nanoscale devices. While bulk GaN has proven to be a highly effective material in numerous applications, its inherent limitations with respect to miniaturization and integration with flexible substrates have prompted intensive research into its 2D counterparts. Notably, theoretical studies have predicted diverse structural phases of 2D GaN, including wurtzite, haeckelite, and hexagonal structures, depending on the number of layers and stacking order [29]. The findings emphasize the material's structural adaptability and the emergence of novel properties distinct from bulk GaN. This structural polymorphism offers a promising avenue for tuning the electronic and optical properties of GaN for advanced device applications. However, the dynamic interplay between interlayer stacking and electronic/optical properties during sliding remains poorly understood.

This study addresses this critical gap by investigating the dynamic modulation of electronic and optical properties in GaN bilayers induced by interlayer sliding. Focusing on hexagonal bilayer GaN, we consider three representative stacking configurations—AA', AB', and AC'—that naturally arise without rotational disorder. Using first-principles density functional theory (DFT) calculations, we systematically analyze the evolution of electronic band structures and optical absorption spectra as a function of interlayer sliding. This approach enables the discernment of the intricate relationship between interlayer interactions, electronic structure, and optical response, thereby providing insights into the dynamic tunability of 2D GaN. Unlike previous studies that primarily focused on static stacking configurations [30, 31], our work explores the dynamic transitions between these configurations, thereby offering a more comprehensive understanding of the structure-property relationships in 2D GaN bilayers. Previous experimental studies have demonstrated that stacking variations in 2D materials can be observed using techniques such as transmission electron microscopy (TEM) and Piezoresponse Force Microscopy (PFM) [32, 33]. In particular, stacking faults in GaN have been investigated with experimental studies with TEM [34]. These studies suggest that interlayer sliding mechanisms may be involved in modifying properties of 2D materials. The findings provide valuable guidelines for the design and optimization of future nanoscale photonic and electronic devices based on 2D GaN, where precise control over interlayer interactions and stacking is important.

## 2. Computational Details

We investigated the interlayer interactions of GaN bilayers using first-principles calculations based on density functional theory (DFT), as implemented in the Vienna *Ab initio* Simulation Package (VASP) [35-37]. The exchange-correlation functional was treated within the generalized gradient approximation (GGA) using the Perdew-Burke-Ernzerhof (PBE) parameterization [38]. Core-valence interactions were described by the projector augmented-wave (PAW) method [37, 39], with a plane-wave basis set and a kinetic energy cutoff of 650 eV. Structural relaxations were performed until residual forces on all atoms were less than 0.01 eV/Å and the total energy converged to within $10^{-6}$ eV. To minimize spurious interactions between periodic images in the non-periodic *z*-direction, a vacuum spacing of at least 15 Å was employed. A 15×15×1 *k*-point mesh was used for structural relaxations and electronic structure calculations, while a denser 30×30×1 mesh was employed for optical property calculations. Interlayer van der Waals interactions were accounted for using Grimme's DFT-D3 correction [40, 41]. The VASPKIT toolkit [42] was used for post-processing and analysis of the calculated data.

The structural stability of the GaN monolayer and bilayers was validated by *ab initio* molecular dynamics (AIMD) simulations (Figure S1, Supplementary Materials). Figure S1 shows the fluctuation of total energy and atomic displacements over a 5 ps simulation time at different temperatures, demonstrating the dynamic stability of the structures. *NVT* ensemble simulations using the Nose-Hoover thermostat to control the temperature were performed at 300 K, 400 K, and 500 K with a 1 fs time step and a total simulation time of 5 ps (5000 steps). The observed stability of the total energy profiles over the temperature range confirms the suitability of these structures for further DFT analysis.

# 3. Results and Discussion

3.1 Structural properties

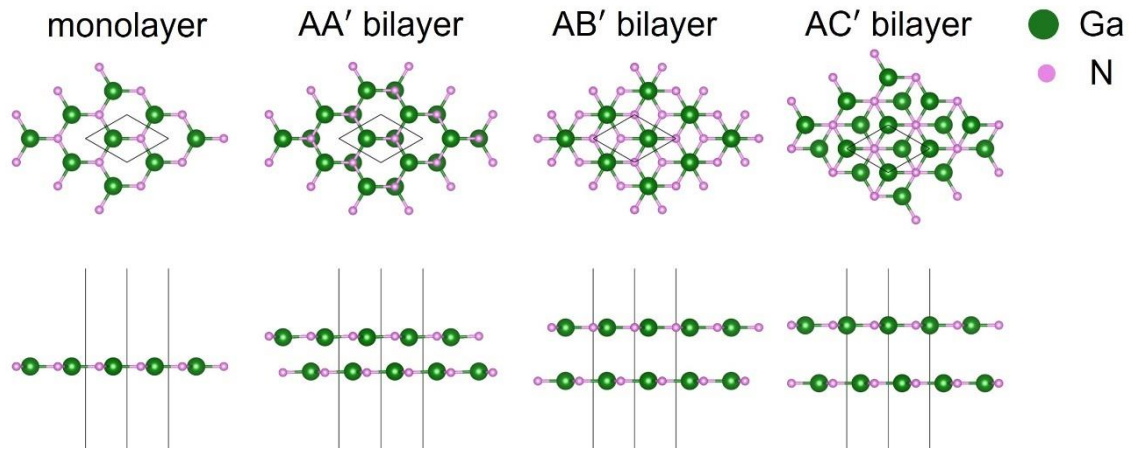

**Figure 1. Top and side views of a monolayer and AA', AB', and AC' stacked GaN bilayers. Ga and N atoms are represented by green and pink spheres, respectively. Unit cells are outlined by black lines.**

We first investigated the structural properties of monolayer and bilayer GaN. Figure 1 shows the top and side views of the considered structures: monolayer, AA', AB', and AC' bilayers. All structures adopt a planar hexagonal lattice similar to that of hexagonal boron nitride (h-BN). The AA' stacking configuration consists of two perfectly aligned hexagonal layers, where atoms of the same species in the top and bottom layers are directly superimposed. This AA' stacking is found to be the most energetically favorable configuration. In the AB' stacking, the Ga atoms of the two layers are vertically aligned, resulting in the N atoms occupying the hollow sites of the opposing layer. In contrast, in AC' stacking, the N atoms are vertically aligned, with the Ga atoms occupying the hollow sites of the opposing layer.

**Table 1. Structural properties of the GaN monolayer and bilayers.**

|  | in-plane lattice constant (Å) | bond length (Å) | bond angle (°) | interlayer distance (Å) |
|---|---|---|---|---|
| monolayer | 3.206 | 1.851 | 120.0 | - |
| AA' bilayer | 3.266 | 1.887 | 119.9 | 2.394 |
| AB' bilayer | 3.204 | 1.850 | 120.0 | 3.594 |
| AC' bilayer | 3.204 | 1.850 | 120.0 | 3.913 |

Table 1 summarizes the key structural parameters of the considered GaN configurations. The in-plane lattice constant of the monolayer is calculated to be 3.206 Å. In the AA'-stacked bilayer, the in-plane lattice constant expands to 3.266 Å, indicating a measurable influence of interlayer interactions. The results are in good agreement with previous studies [43-46]. This expansion is accompanied by an increase in the in-plane Ga-N bond length from 1.851 Å in the monolayer to 1.887 Å in the AA' bilayer. The bond angle remains nearly unchanged, with a value of 119.9° in the AA' bilayer compared to 120.0° in the monolayer. The relatively short interlayer distance in the AA' configuration (2.394 Å) suggests a significant interlayer interaction, stronger than typical van der Waals interactions, approaching a weak chemical bonding regime. This strong interlayer coupling is also manifested in the observed structural buckling of the AA' bilayer, with an out-of-plane displacement of 6.9 pm between the Ga and N atoms within each layer. This buckling is a direct consequence of the interlayer interactions and distinguishes the AA' bilayer from the planar monolayer structure.

In contrast, the AB' and AC' bilayers exhibit significantly larger interlayer distances of 3.594 Å and 3.913 Å, respectively. These larger separations suggest weaker interlayer interactions, which is reflected in their in-plane lattice constants (3.204 Å) and Ga-N bond lengths (1.850 Å) being much closer to those of the monolayer. This indicates that interlayer interactions have a negligible effect on the in-plane structural parameters in the configurations, likely due to the increased interlayer spacing. Furthermore, the observed trend indicates that as the interlayer separation increases and the interlayer coupling weakens, the structural parameters of the bilayer converge to those of an isolated monolayer.

The difference in interlayer spacing between the AB' and AC' stackings is due to the interplay of electrostatic repulsion resulting from the specific atomic arrangements. In the AC' stacking, the nitrogen atoms face each other across the layers. Since nitrogen is highly electronegative, the close proximity of these negatively charged nitrogen atoms results in significant electrostatic repulsion. This repulsion pushes the layers further apart, resulting in a larger interlayer distance. Conversely, in AB' stacking, gallium atoms face each other. Gallium is less electronegative than nitrogen, meaning that the electrostatic repulsion between gallium atoms is weaker than that between nitrogen atoms in the AC' stacking.

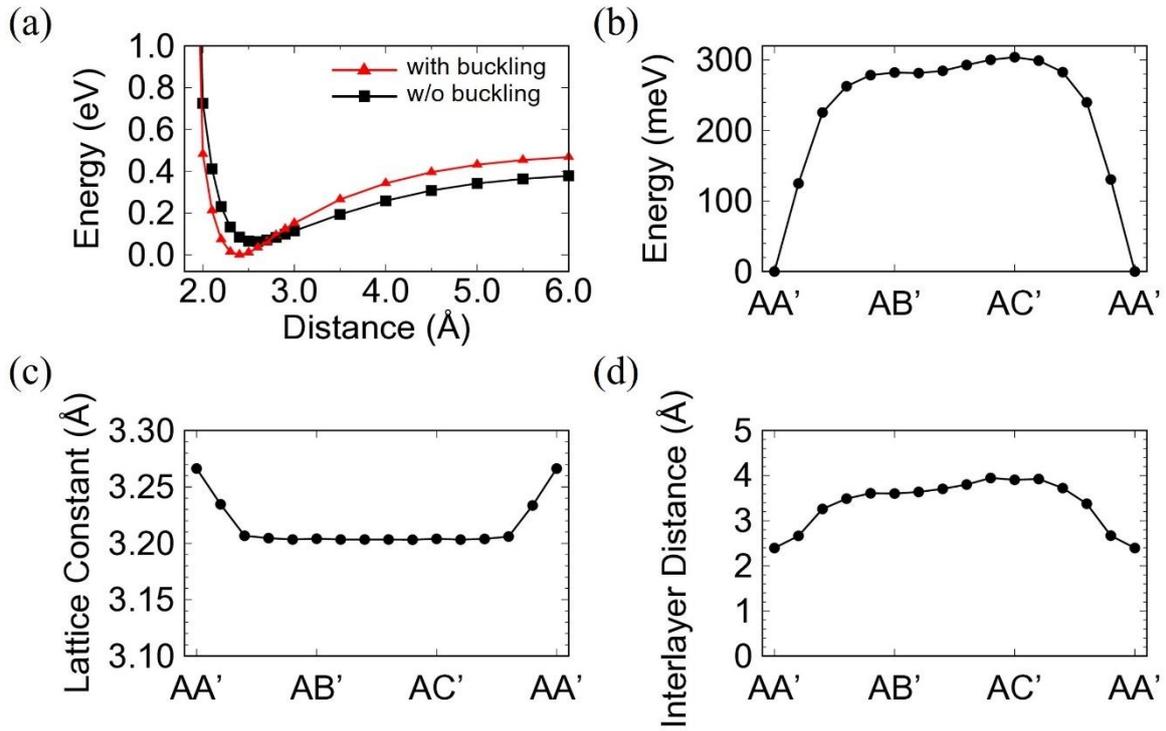

**Figure 2. (a) Total energy per cell of the AA' bilayer as a function of interlayer distance, with and without buckling. (b) Relative energy per cell, (c) in-plane lattice constants, (d) interlayer distances during the sliding process.**

Next, we analyzed the energetics and structural evolution during interlayer sliding. Figure 2(a) shows the total energy of the AA' bilayer as a function of interlayer distance, comparing buckled and unbuckled configurations. As the interlayer distance decreases from large separations, the total energy decreases due to increasing attractive interlayer interactions, primarily van der Waals forces. However, at short interlayer distances, strong repulsive forces arise due to Pauli repulsion, leading to a rapid increase in energy. The buckled configuration reaches its minimum energy at an interlayer distance of 2.4 Å, which is 64 meV lower than the minimum energy of the unbuckled configuration at 2.6 Å. This energetic preference for the buckled structure suggests that the buckling stabilizes the bilayer by optimizing the interlayer bonding environment. The crossing of the two energy curves at ~2.7-2.8 Å indicates a transition from a buckled to an unbuckled ground state. Beyond this distance, the unbuckled structure becomes energetically more favorable, asymptotically approaching the energy of two independent monolayers as interlayer interactions vanish.

Figure 2(b) presents the energy profile during the sliding transitions from the AA' stacking to the AB' and AC' stackings, with the energy of the AA' configuration set to zero. The AB' and AC' stackings are found to be 282 meV and 304 meV higher in energy than the AA' stacking, respectively, indicating their relative instability. The steep energy gradients observed during the AA' → AB' and AC' → AA' transitions suggest significant energy barriers and a strong driving force for the system to relax back to the AA' configuration. In contrast, the relatively gentle slope of the AB' → AC' transition implies a lower energy barrier for this pathway. Notably, a local minimum, approximately 1 meV lower in energy than the AB' configuration, is observed at a displacement slightly beyond the AB' stacking. This local minimum may correspond to a metastable intermediate state during the AB' → AC' transition.

The changes in in-plane lattice constants and interlayer distances during these transitions are shown in Figures 2(c) and 2(d), respectively. The in-plane lattice constant exhibits a maximum value at the AA' stacking and its neighboring points, with a variation of approximately 6.3 pm over all configurations. The interlayer distance increases monotonically from the AA' (2.394 Å) to the AB' (3.602 Å) and AC' (3.907 Å) stackings, which is in good agreement with the weakening of interlayer interactions and the increasing energetic instability. The influence of the structural changes on the electronic properties, including the electronic band structure and partial density of states (PDOS), will be discussed in the following section.

## 3.2 Electronic structures

We now turn to the electronic properties of the 2D GaN structures, analyzing both the evolution of the band structure with varying interlayer spacing and the changes induced by interlayer sliding. Figure 3 presents the electronic band structures and density of states (DOS) for the monolayer and the three bilayer stacking configurations (AA', AB', and AC'). The Fermi level is set to 0 eV as the reference energy.

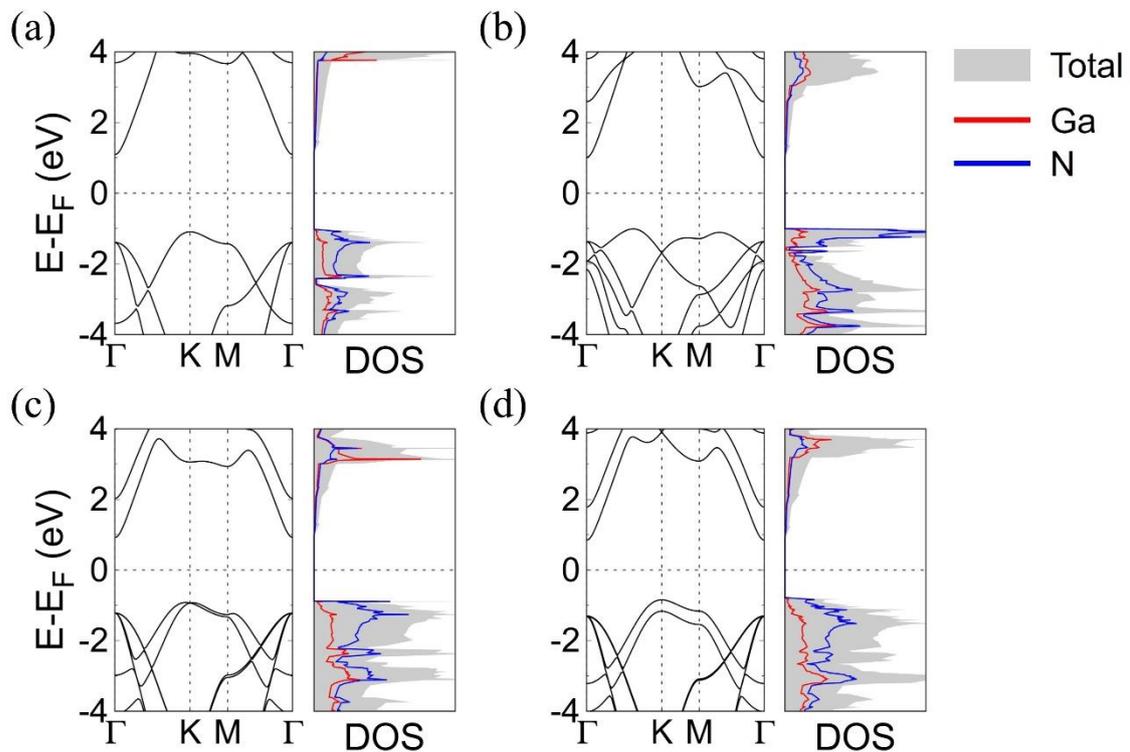

**Figure 3. Electronic band structures (left panels) and density of states (DOS; right panels) for (a) monolayer, (b) AA' bilayer, (c) AB' bilayer, and (d) AC' bilayer GaN. The Fermi level ($E_F$) is set to 0 eV. In the DOS plots, the gray shaded area represents the total DOS, while red and blue lines represent the projected DOS (PDOS) for Ga and N atoms, respectively.**

The calculated indirect band gaps for the monolayer, AA', AB', and AC' bilayers are 2.18 eV, 2.02 eV, 1.84 eV, and 1.69 eV, respectively. The band gaps are within a reasonable range of variation compared to previous studies [47, 48]. The decrease in band gap upon bilayer formation is a direct consequence of interlayer coupling, which leads to splitting of both the valence and conduction bands. This splitting results from the interaction between the electronic states of the two layers, resulting in bonding and antibonding states. The energy difference between the bonding and antibonding states determines the change in band gap. The magnitude of this splitting, and the band gap value, is further modulated by the specific stacking arrangement. In the monolayer in Figure 3(a), the highest valence band, composed mainly of nitrogen p-orbitals, exhibits a maximum at the K point and shows a relatively flat trend between the K and M points. For the AA'-stacked bilayer in Figure 3(b), two bands cross at the K point in this energy range. Also, the band dispersion appears flatter in the regions excluding the K point. This dispersion is directly reflected in the DOS plot, where the nitrogen derived peak near the VBM is significantly sharper and more pronounced compared to the monolayer and other bilayer configurations. This sharper peak suggests an enhanced localization of electronic states due to stronger interlayer coupling in the AA' stacking, where the interlayer distance is much smaller compared to other bilayer configurations. The flattening of the VBM in the AA' bilayer, reflected in the sharp and pronounced peak near the VBM in the DOS, indicates a localization of the hole states. This localization implies an increase in the effective mass of a hole and consequently a reduction in hole mobility, which is expected to negatively affect the hole conductivity of the material.

The electronic structures of the AB' and AC' bilayers, shown in Figure 3(c) and (d), are generally similar to the monolayer, indicating weaker interlayer coupling compared to the AA' configuration. In both the AB' and AC' configurations, the uppermost valence band splits into two. Unlike the AA' configuration, both AB' and AC' configurations retain characteristics similar to the monolayer, with a valence band maximum at the K point and a relatively flat dispersion in the K–M region. In the AB' and AC' bilayers, the N contribution at the VBM is more spread out in the PDOS, indicating weaker interlayer orbital hybridization. This can be attributed to the larger interlayer distances of 3.6 Å in AB' and 3.9 Å in AC', which significantly reduce interlayer coupling compared to the AA' bilayer. Unlike in AA', the broader distribution of nitrogen states in AB' and AC' suggests that the electronic states are more delocalized. A key difference between AB' and AC' lies in the relative positions of these two split bands. In the AB' stacking, the topmost two valence bands cross at K points. This crossing is likely due to a specific combination of interlayer and intralayer orbital interactions resulting from the AB' stacking, leading to a different hybridization pattern compared to the AA' stacking. Moreover, it is also observed that the two bands are nearly degenerate along the K–M path. In contrast, the AC' stacking shows a more uniform separation between the two bands, with no crossing observed. This difference can be attributed to the different relative positions of the atoms in the two layers, which affects the overlap between the orbitals and thus the strength of the interaction.

In addition, a notable feature observed in both AB' and AC' bilayers, and different from the monolayer, is the downward shift of the band just above the lowest conduction band at the Γ point. This band shifts down to about +2 eV (relative to the Fermi level) in both AB' and AC', whereas it is at higher energies in the monolayer. This shift is probably a consequence of the interlayer interaction, although it is weaker than in the AA' configuration. This change in the conduction band structure could also affect the electron transport properties.

The DOS plots of the AB' and AC' bilayers show a doubling of the overall DOS compared to the monolayer, as expected for a bilayer system. However, subtle differences are observed in the PDOS. In the AB' stacking, the Ga peak near +3 eV is sharper and narrower compared to the AC' stacking. This indicates a higher density of Ga-related states at this energy and, importantly, a tendency to localize the states. This localization implies an increase in the electron effective mass and consequently a decrease in the electron mobility at this energy.

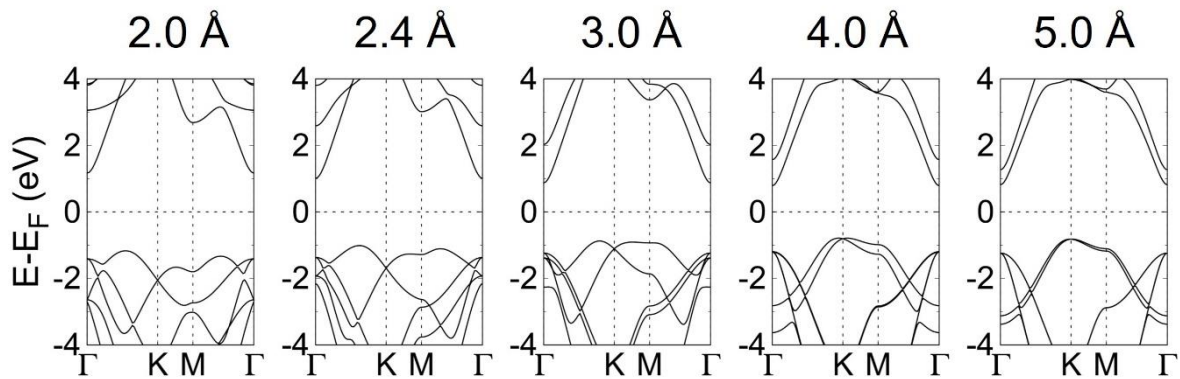

**Figure 4. Electronic band structures of AA'-stacked GaN bilayers as a function of interlayer distance. The interlayer distance (in Å) is indicated above each band structure. The equilibrium interlayer distance is 2.4 Å.**

We then studied the evolution of the topmost valence band in the AA' bilayer as a function of interlayer separation. At large separations, the bilayer effectively behaves as two independent monolayers, exhibiting a band structure characteristic of isolated layers. On the other hand, as the layers approach each other, the increasing interlayer coupling modifies the band structure, especially at the K point of the VBM.

Figure 4 shows the evolution of the electronic band structure of the AA'-stacked bilayer as a function of interlayer distance. At a large spacing of 5.0 Å, the band structure closely resembles that of the monolayer. Systematic changes occur as the interlayer spacing decreases. The most noticeable change occurs in the two highest valence bands. As the interlayer spacing decreases, the crossing point at the K point gradually shifts to lower energy levels. When the interlayer spacing is 5 Å, the crossing point at the K point is located at approximately -1 eV. As the spacing decreases to 2 Å, this crossing point shifts further down to around -2 eV. When the interlayer distance is reduced below the equilibrium spacing of 2.4 Å to 2.0 Å, the highest valence band becomes more distorted, particularly near the M point. This behavior suggests that at such short interlayer distances, strong interatomic repulsion dominates, leading to significant changes in orbital hybridization and the formation of distinct electronic states.

At the same time, the dispersion of the two bands is interesting. At large interlayer separations (e.g., 5.0 Å), the two topmost valence bands show a tendency toward degeneracy. This near-degeneracy at larger separations is a consequence of the weakened interlayer coupling. As the layers become further apart, their electronic states become less perturbed by each other and converge toward the independent monolayer limit. Therefore, the near-merging of the bands at 5.0 Å indicates a decoupling of the layers. As the interlayer distance decreases from 5.0 Å towards the equilibrium value, the crossing at the K point is maintained while the band moves further away, indicating a strengthening of the interlayer interaction. This increased splitting between the bands reflects the formation of distinct bonding and antibonding states due to the enhanced interlayer coupling.

The band structure also evolves during interlayer sliding between different stacking configurations (AA'→AB'→AC'→AA' in Figure S2). These pathways were systematically chosen to include all possible symmetric stacking structures and their intermediate configurations without rotation. As shown in Figure 3, significant differences in the band structures are observed during the AA'→AB' and AC'→AA' transitions. Figure S2 shows how the flat band characteristic of the AA' configuration is modified during these transitions. These results highlight how sliding transitions can significantly alter the band characters, particularly in the valence band region. Moreover, the band gap change during the transitions is shown in Figure S3. These changes in the band gap indicate that interlayer sliding significantly influences the electronic properties. It also suggests that certain stacking arrangements may have distinct electronic characteristics.

3.3 Optical properties

We now turn to the optical properties of 2D GaN. Experimental studies have shown that $I_2$ stacking faults in GaN affect its optical behavior by introducing structural variations that modify its electronic and optical properties [34, 49]. Moreover, research has been conducted on tuning the optical properties of 2D materials, including $MoS_2$ and h-BN through various methods [50, 51]. These support the significance of our theoretical calculations.

The dielectric function, which describes the interaction of the material with light, is critical to understand its optical behavior. The dielectric function is a complex quantity, expressed as:

$$\varepsilon(\omega) = \varepsilon_1 + i\varepsilon_2 \qquad (1)$$

where $\varepsilon_1$ and $\varepsilon_2$ represent the real and imaginary components, respectively. The in-plane symmetry of the 2D GaN structures leads to a diagonal dielectric tensor with equal in-plane components ($\varepsilon_{xx} = \varepsilon_{yy}$) and zero off-diagonal components. Consequently, the optical response in the plane is isotropic. We thus focus on the real and imaginary parts of $\varepsilon_{xx}$ in Figure 5(a) and $\varepsilon_{zz}$ in Figure 5(b), respectively.

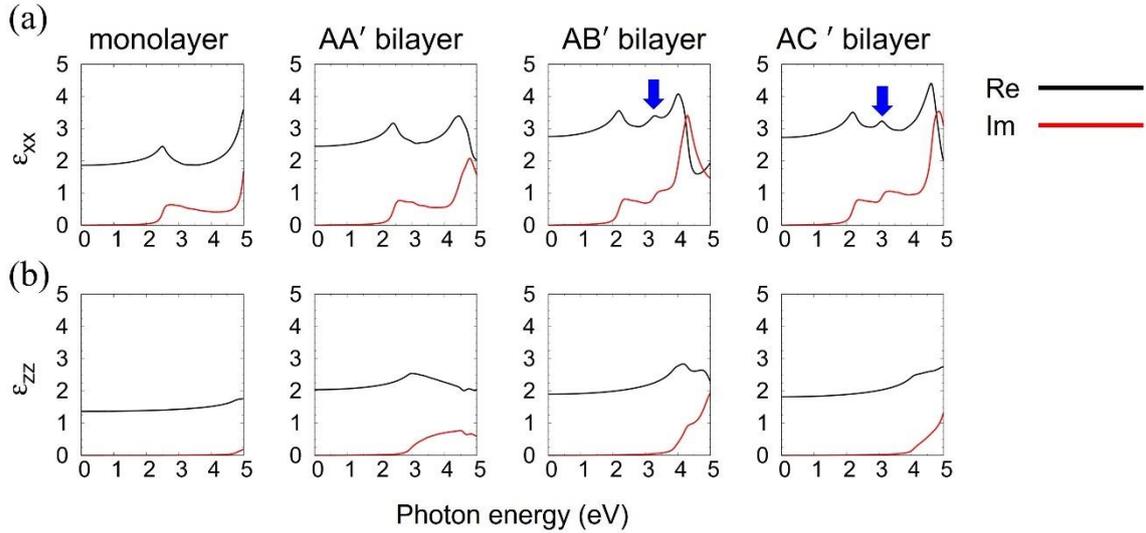

**Figure 5. (a) In-plane ($\varepsilon_{xx}$) and (b) out-of-plane ($\varepsilon_{zz}$) components of frequency-dependent dielectric functions of GaN monolayer, AA', AB', AC' stacking bilayers. The black and red solid lines represent the real and imaginary parts of the function, respectively.**

Figures 5(a) and 5(b) shows the frequency-dependent dielectric functions for the monolayer and various bilayer configurations, plotted as a function of photon energy for both the in-plane ($\varepsilon_{xx}$) and out-of-plane ($\varepsilon_{zz}$) components, respectively. Each graph presents the real ($\varepsilon_1$, black line) and imaginary ($\varepsilon_2$, red line) parts of the dielectric function. We first focus on the in-plane dielectric function ($\varepsilon_{xx}$) shown in Figure 5(a).

Looking at $\varepsilon_1$, the monolayer exhibits a prominent peak around 2.5 eV, followed by a relatively flat region before a sharp increase at ~5.0 eV. This indicates that electronic transitions in the monolayer are concentrated in certain energy regions, with a significant transition occurring at higher energies. In the AA' stacking bilayer, the first peak is observed at about 2.4 eV, close to that of the monolayer. However, there is a significant difference in the position of the second peak, which is shifted to approximately 4.4 eV, occurring at a lower energy than the monolayer peak at ~5.0 eV. This shift implies that the interlayer interaction in the AA' stacking significantly modifies the electronic structure, leading to electronic transitions at different energies.

The AB' and AC' stacking bilayers exhibit three distinct peaks/humps in $\varepsilon_1$, unlike the two peaks observed in the monolayer and AA' stacking. The first peaks for the AB' (~2.2 eV) and AC' (~2.2 eV) configurations are located near the first peaks of the monolayer and AA' configurations. Notably, both AB' and AC' exhibit a second hump/shoulder, indicated by arrows in Figure 5(a), that is absent in the monolayer and AA' stacking. This second hump/shoulder is located at ~3.3 eV for AB' and ~3.1 eV for AC'. The presence of this additional hump/shoulder suggests more complex interlayer interactions in the configurations, enabling additional electronic transitions. The third peaks in the AB' (~4.0 eV) and AC' (~4.6 eV) stackings are also shifted to lower energies compared to the second peak of the monolayer at ~5.0 eV.

Turning to the imaginary part ($\varepsilon_2$) of $\varepsilon_{xx}$ (red line) in Figure 5(a), we observe similar trends across the configurations, although the features are less distinct than in the real part. In the monolayer, the main absorption peak occurs at ~ 5.1 eV, indicating strong optical absorption at this energy. The AA' stacking bilayer exhibits a lower energy peak at ~4.8 eV, which corresponds to the shifted peak observed in the real part ($\varepsilon_1$), further confirming the influence of interlayer coupling. In the AB' and AC' configurations, the highest absorption peaks are observed at ~4.3 eV and ~4.9 eV, respectively. Similar to the real part, both AB' and AC' exhibit an additional feature ~3.4 eV for AB' and ~3.3 eV for AC', which can be described as a hump/shoulder and is absent in the monolayer and AA' stacking. The extra features suggest that the AB' and AC' stackings introduce new electronic transitions, likely due to their distinct interlayer interactions and resulting asymmetry.

Figure 5(b) shows the out-of-plane component ($\varepsilon_{zz}$) of the dielectric function. The monolayer exhibits a nearly flat response, particularly in the real part ($\varepsilon_1$), with minimal variation across the photon energy range. This behavior is expected for a monolayer material, where the electronic structure is predominantly confined within the 2D plane. Since there is no interlayer coupling in a monolayer, electronic transitions polarized along the out-of-plane direction are significantly suppressed, resulting in a nearly constant dielectric response. In contrast, the bilayer structures show more pronounced variation in the $\varepsilon_{zz}$ component, especially above 3 eV in the ultraviolet region. This increased response is attributed to the introduction of interlayer coupling, which enables additional electronic transitions polarized along the out-of-plane direction that are absent in the monolayer.

Interestingly, the imaginary part ($\varepsilon_2$) of $\varepsilon_{zz}$ exhibits distinct behavior among the bilayer stackings. While the $\varepsilon_2$ curve of the AA' stacking appears to saturate around 4-5 eV, the $\varepsilon_2$ curve of the AB' and AC' stackings continues to increase in this energy range. This difference indicates that the AB' and AC' stackings possess additional out-of-plane electronic transitions at higher energies compared to the AA' stacking, likely due to their different interlayer configurations and resulting changes in the out-of-plane electronic structure. This increased out-of-plane response in AB' and AC' contributes to the overall optical anisotropy of these bilayers. The optical properties including 2D conductivity, optical absorptivity, transmissivity, and reflectivity are obtained from the dielectric functions and shown in Figure S4. The bilayers exhibit redshift in the main optical conductivity peaks compared to the monolayer. Also, transmissivity remains high in the infrared region but decreases with increasing photon energy, leading to a rise in absorptivity, while reflectivity remains consistently low.

To further understand the origin of the observed optical features, we calculated the transition dipole moments (TDMs). The sum of the squares of the TDMs provides a measure of the transition probabilities between valence and conduction bands. Our calculations reveal that the most significant transitions occur near the $\Gamma$ point. Comparing the bilayer configurations, we find that the squared TDM for the transition from the valence band at $\Gamma$ to the conduction band at $\Gamma$ is largest in the AA' bilayer. This strong transition explains the prominent first peak observed in the in-plane dielectric function ($\varepsilon_{xx}$) of the AA' bilayer, as shown in Figure 5(a). In the AB' and AC' structures, this transition from the VBM to the CBM at $\Gamma$ also exhibits a large squared TDM, contributing to their respective first peaks. However, a key difference emerges in the AB' and AC' configurations: we observe a significant transition probability from the VBM at $\Gamma$ to the second lowest conduction band at $\Gamma$. This additional transition, which is absent in the monolayer and AA' stacking, accounts for the second hump/shoulder observed in the $\varepsilon_{xx}$ spectra of the AB' and AC' configurations indicated by blue arrows in Figure 5(a). This finding directly links the observed optical features to specific electronic transitions enabled by the distinct interlayer interactions and resulting electronic structures of the AB' and AC' stackings.

The influence of interlayer interactions on the optical properties was also examined. Figure S5 depicts the changes in the in-plane component of the frequency-dependent dielectric function ($\varepsilon_{xx}$) as the interlayer distance is varied relative to the stable AA' configuration. When the interlayer distance is at or below the equilibrium value, the real and imaginary parts of $\varepsilon_{xx}$ exhibit two prominent features, one in the 2-3 eV range and another in the 4-5 eV range. However, as the interlayer spacing increases beyond the equilibrium separation, the higher-energy feature undergoes a splitting, forming two distinct peaks. As the layers are further separated, one of these split peaks shifts progressively towards lower energies, eventually merging with the lower-energy feature near 2 eV at a separation of 5 Å. This evolution of the dielectric function clearly illustrates the significant role of interlayer interactions in determining the optical response, resulting in a pronounced shift of spectral features with varying interlayer distance.

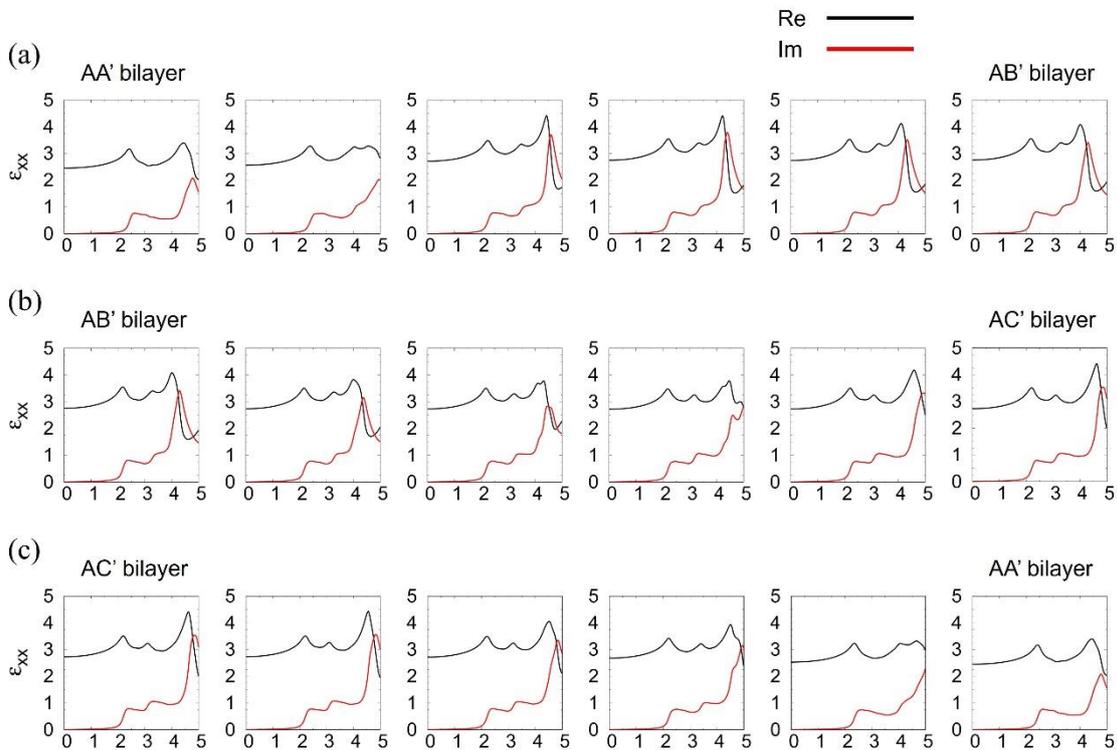

**Figure 6.** The in-plane component of the frequency-dependent dielectric function ($\varepsilon_{xx}$)

as a result of sliding in bilayers. (a) represents sliding process AA' → AB', (b) AB' → AC', and (c) AC' → AA'. The black and red lines indicate the real and imaginary parts, respectively.

Figure 6 presents the evolution of the in-plane dielectric function ($\varepsilon_{xx}$) during interlayer sliding in the bilayers. For the AA' → AB' transition as shown in Figure 6(a), the higher-energy feature around 4-5 eV splits into two distinct peaks. Figure 6(b) depicts the AB' → AC' transition, where the imaginary part exhibits a decrease in peak intensity at the midpoint of the sliding path, followed by a subsequent increase as the configuration approaches AC'. Figure 6(c) shows the AC' → AA' transition, which exhibits a behavior similar to the reverse of the AA' → AB' transition, retracing the changes observed in Figure 6(a).

## 4. Conclusion

In this study, we investigated the structural, electronic, and optical properties of GaN bilayers, focusing on the influence of stacking configurations (AA', AB', and AC') and interlayer spacings and sliding. By analyzing static configurations and intermediate stages during sliding transitions and interlayer spacing variations, we elucidated the intricate interplay between these parameters and the behavior of the material. Our structural analysis revealed that the AA' stacking is the most stable configuration, exhibiting significant interlayer interactions that induce structural buckling, a feature absent in the less stable AB' and AC' stackings. The properties of the AB' and AC' stackings increasingly resemble those of two weakly interacting monolayers, demonstrating the essential role of interlayer coupling in determining bilayer properties. Electronic structure analysis revealed a correlation between structural stability and band gap: less stable stackings exhibit smaller band gaps. In particular, the strong interlayer coupling in the AA' stacking leads to a pronounced change in valence band.

Our optical property analysis further highlighted the sensitivity of GaN optical response to stacking configuration and interlayer distance. The AA' stacking displayed distinct optical signatures. Importantly, the emergence of additional optical transitions in the AB' and AC' configurations, attributed to their distinct interlayer interactions, opens up new possibilities for tailoring the optical response of GaN. This tunability, achieved through the manipulation of stacking arrangements and interlayer spacing, presents significant opportunities for designing advanced photonic and optoelectronic devices. To further refine the optical spectra and explore excitonic effects, future studies could incorporate GW+BSE calculations for a more accurate description of many-body interactions.

This study establishes a comprehensive framework for understanding and exploiting the tunability of GaN bilayers through stacking and interlayer engineering. By demonstrating the profound influence of these parameters on both electronic and optical properties, our findings provide valuable insights into the design of advanced two-dimensional materials and pave the way for the design of GaN-based devices with tailored functionalities, including high-performance transistors, photodetectors, and light-emitting diodes, achieved through precise control over interlayer interactions.

## Acknowledgements

This research was supported by the Basic Science Research Program through the National Research Foundation of Korea (NRF) funded by the Ministry of Education (NRF-2020R1A6A1A03043435).

## References

[1] M. Rais-Zadeh, V.J. Gokhale, A. Ansari, M. Faucher, D. Théron, Y. Cordier, L. Buchaillot, Gallium nitride as an electromechanical material, Journal of Microelectromechanical Systems, 23 (2014) 1252-1271.

[2] R. Schwarz, K. Khachaturyan, E. Weber, Elastic moduli of gallium nitride, Appl. Phys. Lett., 70 (1997) 1122-1124.

[3] S. Nakamura, M.R. Krames, History of gallium–nitride-based light-emitting diodes for illumination, Proc. IEEE, 101 (2013) 2211-2220.

[4] S. Rajbhandari, J.J. McKendry, J. Herrnsdorf, H. Chun, G. Faulkner, H. Haas, I.M. Watson, D. O'Brien, M.D. Dawson, A review of gallium nitride LEDs for multi-gigabit-per-second visible light data communications, Semicond. Sci. Technol., 32 (2017) 023001.

[5] G. Fasol, Room-temperature blue gallium nitride laser diode, Science, 272 (1996) 1751-1752.

[6] T. Melo, Y. Hu, C. Weisbuch, M. Schmidt, A. David, B. Ellis, C. Poblenz, Y. Lin, M. Krames, J. Raring, Gain comparison in polar and nonpolarsemipolar gallium-nitride-based laser diodes, Semicond. Sci. Technol., 27 (2012) 024015.

[7] K. Peng, S. Eskandari, E. Santi, Characterization and modeling of a gallium nitride power

HEMT, IEEE Transactions on Industry Applications, 52 (2016) 4965-4975.

[8] A.A. Fletcher, D. Nirmal, A survey of Gallium Nitride HEMT for RF and high power applications, Superlattices Microstruct., 109 (2017) 519-537.

[9] A. Udabe, I. Baraia-Etxaburu, D.G. Diez, Gallium nitride power devices: a state of the art review, IEEE Access, 11 (2023) 48628-48650.

[10] S. Nakamura, Current status of GaN-based solid-state lighting, MRS Bull., 34 (2009) 101-107.

[11] S. Li, A. Waag, GaN based nanorods for solid state lighting, J. Appl. Phys., 111 (2012).

[12] A. Lidow, J. Strydom, R. Strittmatter, C. Zhou, GaN: A Reliable Future in Power Conversion: Dramatic performance improvements at a lower cost, IEEE Power Electronics Magazine, 2 (2015) 20-26.

[13] A. Lidow, M. De Rooij, J. Strydom, D. Reusch, J. Glaser, GaN transistors for efficient power conversion, John Wiley & Sons, 2019.

[14] W.-S. Choi, J.-Y. Kim, J.-H. Lee, S.-H. Choi, 6G Technology Competitiveness and Network Analysis: Focusing on GaN Integrated Circuit Patent Data, Journal of Industrial Convergence, 21 (2023) 1-15.

[15] T.K. Sahu, S.P. Sahu, K. Hembram, J.-K. Lee, V. Biju, P. Kumar, Free-standing 2D gallium nitride for electronic, excitonic, spintronic, piezoelectric, thermoplastic, and 6G wireless communication applications, NPG Asia Mater., 15 (2023) 49.

[16] I. Guy, S. Muensit, E. Goldys, Extensional piezoelectric coefficients of gallium nitride and aluminum nitride, Appl. Phys. Lett., 75 (1999) 4133-4135.

[17] C.M. Foster, R. Collazo, Z. Sitar, A. Ivanisevic, Aqueous stability of Ga-and N-polar gallium nitride, Langmuir, 29 (2013) 216-220.

[18] K.S. Novoselov, A.K. Geim, S.V. Morozov, D.-e. Jiang, Y. Zhang, S.V. Dubonos, I.V. Grigorieva, A.A. Firsov, Electric field effect in atomically thin carbon films, Science, 306

(2004) 666-669.

[19] G. Cassabois, P. Valvin, B. Gil, Hexagonal boron nitride is an indirect bandgap semiconductor, Nat. Photonics, 10 (2016) 262-266.

[20] K. Zhang, Y. Feng, F. Wang, Z. Yang, J. Wang, Two dimensional hexagonal boron nitride (2D-hBN): synthesis, properties and applications, Journal of Materials Chemistry C, 5 (2017) 11992-12022.

[21] K.K. Kim, H.S. Lee, Y.H. Lee, Synthesis of hexagonal boron nitride heterostructures for 2D van der Waals electronics, Chem. Soc. Rev., 47 (2018) 6342-6369.

[22] Q.H. Wang, K. Kalantar-Zadeh, A. Kis, J.N. Coleman, M.S. Strano, Electronics and optoelectronics of two-dimensional transition metal dichalcogenides, Nat. Nanotechnol., 7 (2012) 699-712.

[23] W. Choi, N. Choudhary, G.H. Han, J. Park, D. Akinwande, Y.H. Lee, Recent development of two-dimensional transition metal dichalcogenides and their applications, Mater. Today, 20 (2017) 116-130.

[24] S. Manzeli, D. Ovchinnikov, D. Pasquier, O.V. Yazyev, A. Kis, 2D transition metal dichalcogenides, Nature Reviews Materials, 2 (2017) 1-15.

[25] R. Yang, Y. Fan, Y. Zhang, L. Mei, R. Zhu, J. Qin, J. Hu, Z. Chen, Y. Hau Ng, D. Voiry, 2D transition metal dichalcogenides for photocatalysis, Angew. Chem., 135 (2023) e202218016.

[26] C. Fong, S. Ng, F. Yam, H. Abu Hassan, Z. Hassan, Synthesis of two-dimensional gallium nitride via spin coating method: influences of nitridation temperatures, J. Sol-Gel Sci. Technol., 68 (2013) 95-101.

[27] Z.Y. Al Balushi, K. Wang, R.K. Ghosh, R.A. Vilá, S.M. Eichfeld, J.D. Caldwell, X. Qin, Y.-C. Lin, P.A. DeSario, G. Stone, Two-dimensional gallium nitride realized via graphene encapsulation, Nat. Mater., 15 (2016) 1166-1171.

[28] N.A. Koratkar, Two-dimensional gallium nitride, Nat. Mater., 15 (2016) 1153-1154.

[29] Y. Jia, Z. Shi, W. Hou, H. Zang, K. Jiang, Y. Chen, S. Zhang, Z. Qi, T. Wu, X. Sun, Elimination of the internal electrostatic field in two-dimensional GaN-based semiconductors, npj 2D Materials and Applications, 4 (2020) 31.

[30] D. Xu, H. He, R. Pandey, S.P. Karna, Stacking and electric field effects in atomically thin layers of GaN, J. Phys.: Condens. Matter, 25 (2013) 345302.

[31] X. Cai, S. Deng, L. Li, L. Hao, A first-principles theoretical study of the electronic and optical properties of twisted bilayer GaN structures, J. Comput. Electron., 19 (2020) 910-916.

[32] J.P. Oviedo, S. KC, N. Lu, J. Wang, K. Cho, R.M. Wallace, M.J. Kim, In situ TEM characterization of shear-stress-induced interlayer sliding in the cross section view of molybdenum disulfide, ACS nano, 9 (2015) 1543-1551.

[33] K. Yasuda, X. Wang, K. Watanabe, T. Taniguchi, P. Jarillo-Herrero, Stacking-engineered ferroelectricity in bilayer boron nitride, Science, 372 (2021) 1458-1462.

[34] A. Lang, J. Hart, J. Wen, D. Miller, D. Meyer, M. Taheri, I2 basal stacking fault as a degradation mechanism in reverse gate-biased AlGaN/GaN HEMTs, Appl. Phys. Lett., 109 (2016).

[35] G. Kresse, J. Furthmüller, Efficiency of ab-initio total energy calculations for metals and semiconductors using a plane-wave basis set, Comput. Mater. Sci., 6 (1996) 15-50.

[36] G. Kresse, J. Furthmüller, Efficient iterative schemes for ab initio total-energy calculations using a plane-wave basis set, Phys. Rev. B, 54 (1996) 11169.

[37] G. Kresse, D. Joubert, From ultrasoft pseudopotentials to the projector augmented-wave method, Phys. Rev. B, 59 (1999) 1758.

[38] J.P. Perdew, K. Burke, M. Ernzerhof, Generalized gradient approximation made simple, Phys. Rev. Lett., 77 (1996) 3865.

[39] P.E. Blöchl, Projector augmented-wave method, Phys. Rev. B, 50 (1994) 17953.

[40] S. Grimme, Semiempirical GGA-type density functional constructed with a long-range dispersion correction, J. Comput. Chem., 27 (2006) 1787-1799.

[41] S. Grimme, J. Antony, S. Ehrlich, H. Krieg, A consistent and accurate ab initio parametrization of density functional dispersion correction (DFT-D) for the 94 elements H-Pu, J. Chem. Phys., 132 (2010).

[42] V. Wang, N. Xu, J.-C. Liu, G. Tang, W.-T. Geng, VASPKIT: A user-friendly interface facilitating high-throughput computing and analysis using VASP code, Comput. Phys. Commun., 267 (2021) 108033.

[43] Y. Shen, H. Liu, Q. Zhang, Y. Zhang, X. Yang, B. Wang, First principles study on toxic gas adsorption of PtSe2/GaN heterostructure modified by Cu-group elements, Appl. Surf. Sci., 670 (2024) 160678.

[44] K. Ren, S. Wang, Y. Luo, Y. Xu, M. Sun, J. Yu, W. Tang, Strain-enhanced properties of van der Waals heterostructure based on blue phosphorus and g-GaN as a visible-light-driven photocatalyst for water splitting, RSC Adv., 9 (2019) 4816-4823.

[45] L. Huang, Q. Yue, J. Kang, Y. Li, J. Li, Tunable band gaps in graphene/GaN van der Waals heterostructures, J. Phys.: Condens. Matter, 26 (2014) 295304.

[46] A.K. Augustin Lu, T. Yayama, T. Morishita, M.J. Spencer, T. Nakanishi, Uncovering new buckled structures of bilayer GaN: A first-principles study, J. Phys. Chem. C, 123 (2018) 1939-1947.

[47] X. Dong, Z. Peng, T. Chen, L. Xu, Z. Ma, G. Liu, K. Cen, Z. Xu, G. Zhou, Electronic structures and transport properties of low-dimensional GaN nanoderivatives: A first-principles study, Appl. Surf. Sci., 561 (2021) 150038.

[48] H. Şahin, S. Cahangirov, M. Topsakal, E. Bekaroglu, E. Akturk, R.T. Senger, S. Ciraci, Monolayer honeycomb structures of group-IV elements and III-V binary compounds: First-principles calculations, Physical Review B—Condensed Matter and Materials Physics, 80

(2009) 155453.

[49] I. Tischer, M. Feneberg, M. Schirra, H. Yacoub, R. Sauer, K. Thonke, T. Wunderer, F. Scholz, L. Dieterle, E. Müller, I 2 basal plane stacking fault in GaN: Origin of the 3.32 eV luminescence band, Physical Review B—Condensed Matter and Materials Physics, 83 (2011) 035314.

[50] B. Liu, Z. Zhang, K. Liao, R. Wu, C. Zhu, H. Xie, C. Zha, Y. Yin, X. Jiang, S. Qin, Tuning optical properties of monolayer MoS2 through the 0D/2D interfacial effect with C60 nanoparticles, Appl. Surf. Sci., 523 (2020) 146371.

[51] H.Y. Lee, M.M. Al Ezzi, N. Raghuvanshi, J.Y. Chung, K. Watanabe, T. Taniguchi, S. Garaj, S. Adam, S. Gradecak, Tunable optical properties of thin films controlled by the interface twist angle, Nano Lett., 21 (2021) 2832-2839.

# Supplementary Materials

# Dynamic Modulation of Electronic and Optical Properties in GaN Bilayers by Interlayer Sliding

Heeju Kim[1] and Gunn Kim[1*]

[1]*Department of Physics & Astronomy and Hybrid Materials Research Center, Sejong University, Seoul 05006, Republic of Korea*

*Corresponding author: gunnkim@sejong.ac.kr

# 1. Thermal stability analysis of GaN using molecular dynamics simulations

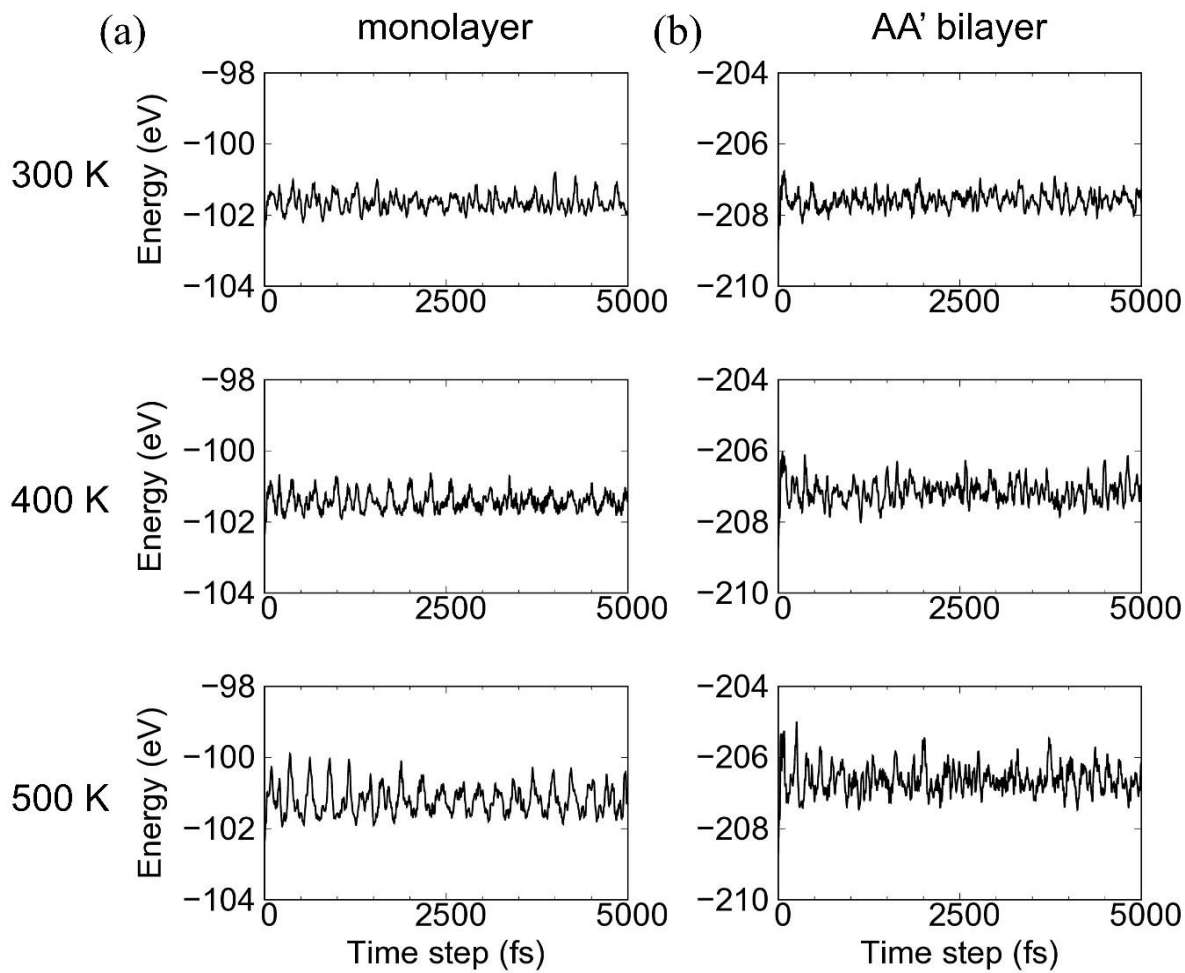

**Figure S1.** Total energy as a function of time during molecular dynamics simulations for (a) monolayer and (b) AA' bilayer systems at 300 K, 400 K, and 500 K.

To investigate the structural stability of two-dimensional hexagonal gallium nitride (2D h-GaN), we performed *ab initio* molecular dynamics (AIMD) simulations. Details of the simulation methods are provided in Section 2. Figure S1 shows the time evolution of the total energy for monolayer and AA'-stacked bilayer h-GaN at temperatures of 300 K, 400 K, and 500 K. In all simulated systems, the total energy exhibited stable fluctuations around a mean value, demonstrating the absence of significant structural transformations or irreversible changes within the simulation timescale. This indicates the intrinsic structural stability of 2D h-GaN under the investigated temperature range.

## 2. Sliding-induced modulation of electronic band structure of GaN bilayer

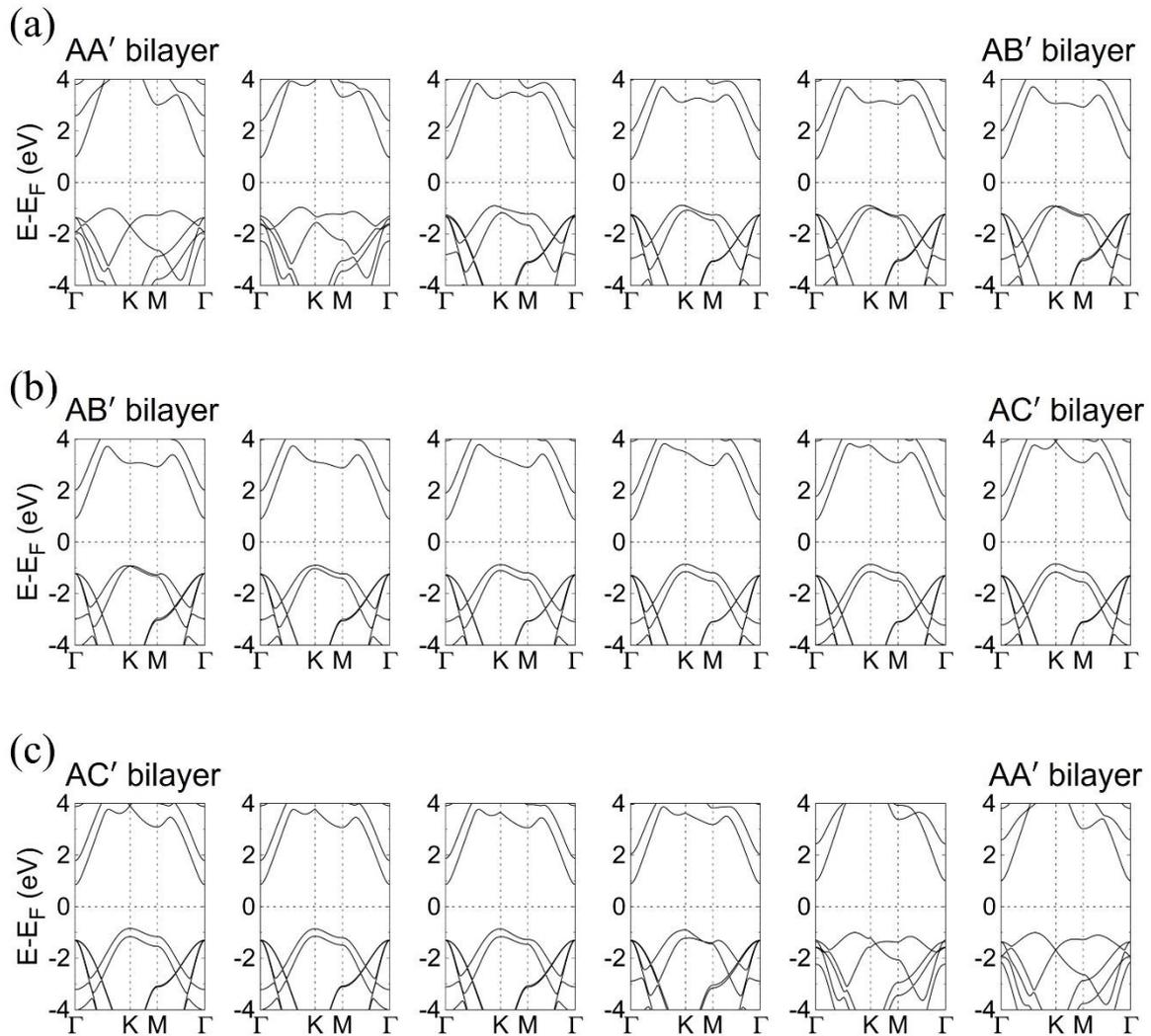

**Figure S2. Electronic band structure changes associated with the sliding transformations: (a) AA' to AB', (b) AB' to AC', and (c) AC' to AA'.**

Figure S2 shows the evolution of the electronic band structure during sliding transitions between symmetric stacking configurations: AA' → AB', AB' → AC', and AC' → AA'. For each transition between symmetric configurations, four intermediate configurations were analyzed, generated by incrementally shifting atoms by about 0.4 Å along the in-plane sliding direction. Significant modifications were observed in the valence band and the band immediately below it. During the AA' → AB' transition in Figure S2(a), the two bands near the valence band maximum gradually converge in energy, especially between K and M points. In this process, the band crossing at the K point disappears in the intermediate asymmetric structures but reappears in the symmetric AB' configuration. During the AB' → AC' transition shown in Figure S2(b), the two bands of interest remain energetically close. However, the band crossings observed at K and M points in the AB' configuration vanish rapidly with even minor deviations from the AB' structure. The AC' → AA' transition in Figure S2(c) shows a distinct behavior: when the structure is closer to AC', the two relevant bands remain fully separated along the K–M path. As the transition progresses towards AA', the bands gradually approach each other along the same path, eventually meeting. Near the end of this transition, a clear band crossing emerges between K and M, ultimately setting at the K point in the fully formed AA' configuration. This analysis reveals the sensitivity of the electronic structure, especially the valence bands, to the interlayer stacking arrangements during sliding transitions, demonstrating how atomic arrangement significantly influences the electronic properties of the system.

## 3. The band gap change during the sliding

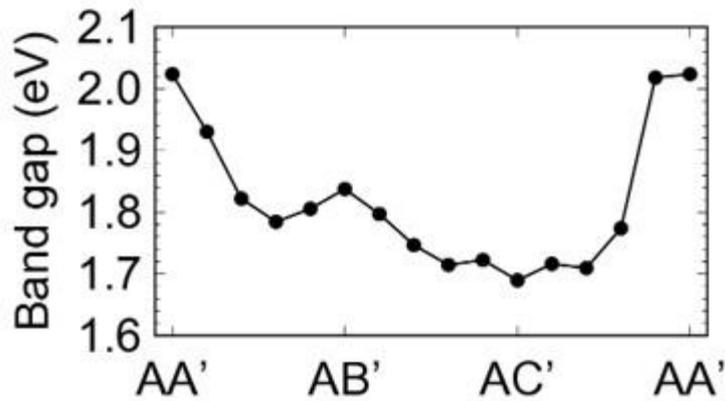

**Figure S3. The band gap change during the sliding process.**

Figure S3 shows the band gap evolution during the interlayer sliding process. Starting from 2.02 eV in the most stable AA' stacking, the bandgap decreases as the system shifts to the AB' stacking configuration, reaching a local minimum of 1.78 eV before rising slightly to 1.84 eV at AB'. Continued sliding to the AC' stacking further reduces the band gap to a minimum value of 1.69 eV. As the system transitions from AC' back to AA', the bandgap increases, returning to the initial value of 2.03 eV. Notably, a region within this process exhibits significant bandgap variation, highlighting an abrupt shift in the electronic structure.

## 4. The optical properties of 2D GaN

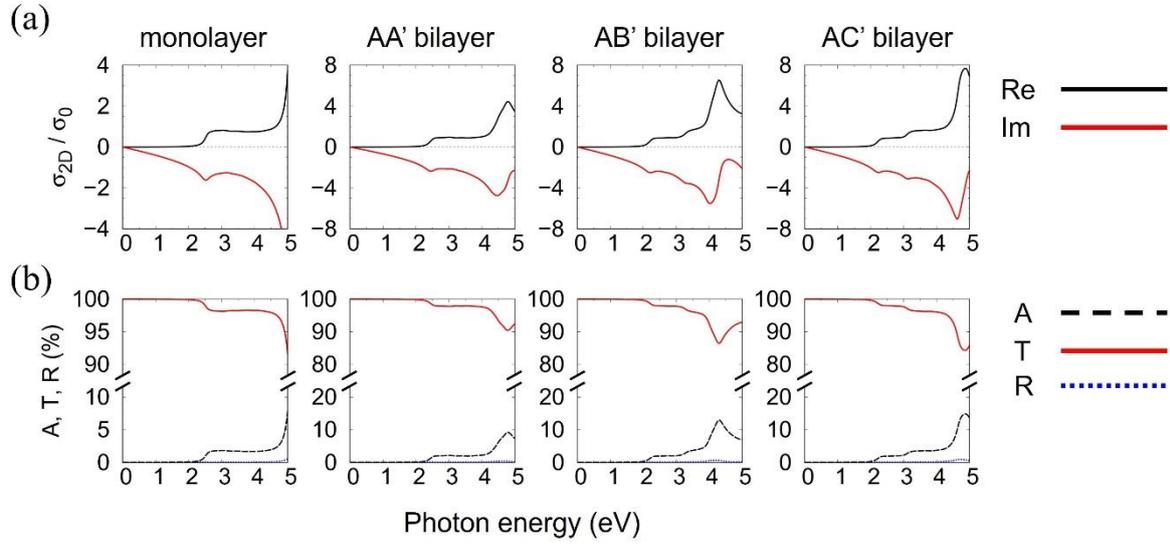

**Figure S4. Optical properties of GaN monolayer and bilayers. (a) 2D optical conductivity with $\sigma_0 = e^2/(4\hbar)$. (b) Optical absorptivity (A), transmissivity (T), and reflectivity (R), with corresponding line styles: dashed black (A), solid red (T), and dotted blue (R).**

We examine the optical conductivity, which can be calculated from the dielectric function. The optical conductivity in 2D materials is expressed using the vacuum layer thickness (L) as follows:

$$\sigma_{2D}(\omega) = iL[1 - \epsilon(\omega)]\epsilon_0 \omega.$$

Here, $\epsilon(\omega)$ represents the complex dielectric function dependent on frequency, $\epsilon_0$ is the permittivity of vacuum, and $\omega$ is the frequency of the incident light.

Figure S4(a) shows the 2D optical conductivity of the GaN monolayer and various bilayer configurations. For the monolayer, the real part of the conductivity (black line) exhibits a significant increase above 4 eV, which lies outside the typical energy window of interest in this figure. The imaginary part (red line) remains negative with a dip near 4 eV. In the bilayers (AA', AB', AC'), the real part of the conductivity remains positive across the energy range. The main peaks observed in the bilayers are shifted to lower energies compared to the monolayer, bringing them within the displayed energy window. In particular, the AA' bilayer exhibits the lowest peak energy, followed by AB', and then AC', indicating a progressive shift to higher energies among the bilayers, but still lower than the monolayer peak. In addition, the peak heights increase in the same order, with AA' showing the smallest peak, followed by AB', and AC' displaying the largest peak. The imaginary part of the conductivity also exhibits distinct trends. While the monolayer shows a dip near 4 eV, the bilayers display similar dips with those increases progressively from AA' to AC'.

Figure S4(b) shows the optical absorptivity (A), transmissivity (T), and reflectivity (R) spectra for the GaN monolayer and bilayers as a function of photon energy. For clarity, the y-axis scales are different for the monolayer and bilayers, and a broken y-axis is employed. Across all structures, the transmissivity is close to 100% in the infrared region. As the photon energy increases into the visible range, the transmissivity gradually decreases, accompanied by a corresponding increase in absorptivity. The reflectivity remains consistently low across all configurations throughout this energy range for all configurations.

## 5. Dependence of the in-plane frequency-dependent dielectric function on interlayer distance

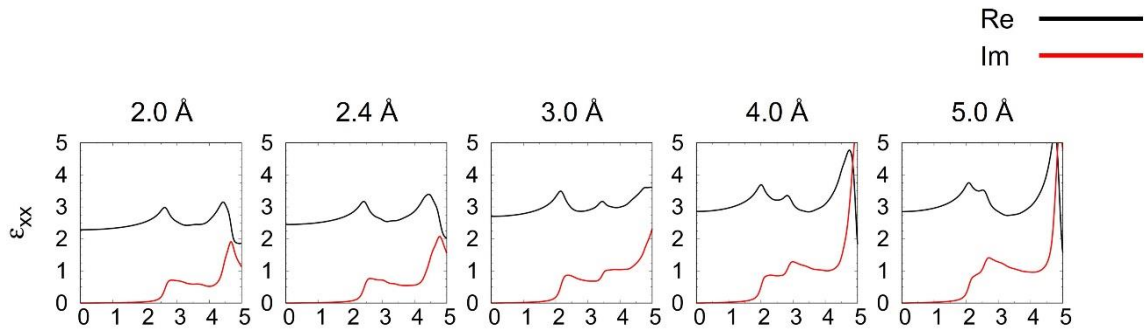

**Figure S5. Frequency-dependent in-plane dielectric function ($\varepsilon_{xx}$) for the AA' stacking bilayer system at varying interlayer distances. The values above each panel indicate the interlayer distance, with the second panel (2.4 Å) corresponding to the stable interlayer spacing.**

Figure S5 shows the evolution of the in-plane component of the frequency-dependent dielectric function ($\varepsilon_{xx}$) as a function of interlayer distance, relative to the equilibrium AA' stacking configuration. At or near the equilibrium interlayer spacing, the real and imaginary parts of the dielectric function exhibit two prominent features: a distinct peak in the 2–3 eV and the 4–5 eV range. As the interlayer spacing increases beyond the equilibrium value, a third distinct peak emerges for the real part of the dielectric function ($\varepsilon_{xx}$). This new peak originates from the higher-energy feature (4–5 eV) observed at equilibrium spacing and undergoes a progressive redshift (shift to lower energies) with increasing interlayer separation. In the case of the imaginary part of the dielectric function ($\varepsilon_{xx}$), a new feature emerges that resembles a step-like structure at certain interlayer separations such as 3 Å or 4 Å. Notably, at an interlayer spacing of ~ 5 Å, this redshifted peak converges with the lower-energy peak in the 2–3 eV range, resulting in significant spectral overlap and a pronounced change in the dielectric response. The findings demonstrate the strong dependence of the in-plane dielectric response on interlayer spacing, revealing the intricate interplay between electronic structure and interatomic distances in this layered material.